\documentclass[mathleft
]{an}
\usepackage{graphicx}
\usepackage{times}
\usepackage{amsmath}
\begin{document}

\Pagespan{789}{}
\Yearpublication{2011}%
\Yearsubmission{2010}%
\Month{11}%
\Volume{999}%
\Issue{88}%

\title{Power calculation for gravitational radiation: oversimplification 
and the importance of time scale}

\author{Alan B.\ Whiting\inst{1}\fnmsep\thanks{Corresponding author:
  \email{abw@star.sr.bham.ac.uk}\newline}
}
\titlerunning{oversimplification}
\authorrunning{A. B. Whiting}
\institute{Astrophysics and Space Research Group, School of Physics and
Astronomy,
University of Birmingham, Edgbaston, Birmingham B15 2TT, UK}

\received{2010 June}
\accepted{2010 September}
\publonline{2010 September 17}


\abstract{A simplified formula for gravitational-radiation power is
examined.  It is shown to give completely erroneous answers in
three situations, making it useless even for rough estimates.  It is
emphasised that short timescales, as well as fast speeds, make
classical approximations to relativistic calculations untenable.}

\maketitle

\section{Introduction}

Gravitational radiation, the vibration of the space-time metric produced by
masses in motion, forms one of the accepted predictions of General 
Relativity.  Though not yet directly detected, current projects
are producing astrophysically interesting upper bounds and 
planned instruments may well make the first real observations\footnote{The
literature on gravitational radiation is vast and even a brief
survey is beyond the scope of this paper.  Basic derivations are
found in most General Relativity texts, and specific references
to Landau \& Lifshitz (1975) will appear below.  Dietz (2010),
to take a recent example, shows the 
useful application
of non-observations to astronomical objects.}.

The calculation of gravitational radiation in all but the simplest
cases can be difficult and tedious, however.  Baker (2006) presented
a simplified formula for the gravitational power radiated by an
object, or collection of objects, `in order to render astrophysical
applications more apparent.' This paper examines that formula to
determine its limits of application.

\section{Baker's derivation}

The starting point for Baker's (2006) derivation is the formula for
total averaged power radiated by a body, under various simplifying
assumptions (the gravitational waves are of small amplitude, the
stress-energy tensor can be reduced to the mass density).
From Landau \& Lifshitz (1975),
p.\ 355, eq.\ 110.16, this is

\begin{equation}
-\frac{\mathrm{d}E}{\mathrm{d}t} = P = \frac{G}{45c^5} 
\left| \dddot{\mathrm{D}}_{\alpha \beta} \right|^2
\label{power}
\end{equation}
(I have used $G$ for the gravitational constant instead of Landau \&
Lifshitz' $k$), where the moment of inertia tensor is given by

\begin{equation}
\mathrm{D}_{\alpha\beta} = \int \rho \left( 3 x_{\alpha} x_{\beta}
- \delta_{\alpha \beta} r^2 \right) \mathrm{d}V,
\label{inertia}
\end{equation}
the integral to be taken over the volume of the body in question, and
transformed into a sum when considering a collection of point masses.
(This is their equation 110.10, p.\ 355; 
I have substituted $\rho$ for the mass density instead of Landau \&
Lifshitz' $\mu$, since the latter symbol appears with a different
meaning below.)

Citing considerations of `symmetry' (which he does not specify), Baker
equates $\mathrm{D}_{\alpha \beta}$ with a scalar moment of inertia
$I$, taken to be a mass $\delta m$ times the square of the radius of
gyration $r$.  In taking the third time derivative he obtains
\begin{equation}
\frac{\mathrm{d}^3 I}{\mathrm{d} t^3}=2r \delta m \left( 
\frac{\mathrm{d}^3 r}{\mathrm{d} t^3} \right) + \dots
\end{equation}
(Baker does not say what the missing terms on the right are, nor why he
chose to ignore them).  He then identifies $\delta m$ times the third
time derivative of $r$ with the time derivative of force; and finally
averages the change in force over some time interval, giving as his 
formula for power

\begin{equation}
P = 1.76 \times 10^{-52} \left( 2r \frac{\Delta f_t}{\Delta t}
\right)^2
\label{Baker}
\end{equation}
the numerical coefficient chosen to give watts\footnote{I have not been
able to duplicate Baker's numbers, which he gives in two incompatible
equations.  In all numerical calculations I will use $6.1 \times
10^{-55}$s$^3$ kg$^{-1}$ m$^{-2}$ as the coefficient to the moment-of-inertia
term in Eq.\,(\ref{power}); the difference, large as it is, does
not affect my conclusions.}.  He then applies his formula to two-body
motion and rigid-rod rotation, finding numerical agreement, and considers
his formula to have general application.

\section{Two testing situations}

A thorough investigation of Eq.\,(\ref{Baker}) would set out the conditions
under which the simplifications and assumptions of its derivation hold
good.  For present purposes it is sufficient to look at three examples.

First, we follow on from Problem 1, page 356 of Landau \& Lifshitz (1975).
Two masses, $m_1$ and $m_2$ are in a circular orbit a distance $r$
apart.  They move with angular frequency $\omega$ and at time zero
lie along the $x$-axis, rotating in the $x, y$ plane.  The components
of the moment of inertia tensor are

\begin{eqnarray}
\mathrm{D}_{xx} & = & \mu r^2 \left( 3 \cos^2 (\omega t) - 1 \right) \\
\nonumber
& = & \frac{3}{2} \mu r^2 \cos (2 \omega t ) + \frac{1}{2} \mu r^2 \\ \nonumber
\mathrm{D}_{yy} & = & \mu r^2 \left( 3 \sin^2 (\omega t) - 1 \right) \\
\nonumber
& = &\frac{1}{2} \mu r^2 - \frac{3}{2} \mu r^2 \cos (2 \omega t ) \\ \nonumber
\mathrm{D}_{xy} & = & \mu r^2 \left( 3 \cos (\omega t) \sin (\omega t) 
\right) \\ \nonumber
& = & \frac{3}{2} \mu r^2 \sin (2 \omega t) \\ \nonumber
\mathrm{D}_{zz} & = & - \mu r^2
\end{eqnarray}
where $\mu = m_1 m_2/(m_1 + m_2)$ is the reduced mass.  
Next, we add an identical pair of
masses, the same distance apart rotating in the same orbit with the
same speed, but place them one-quarter of the way around the orbit
with respect to the first pair.  They add the following terms to
the moment of inertia tensor:
\begin{eqnarray}
\mathrm{D}_{xx}' & = & \frac{3}{2} \mu r^2 \cos (2 \left[
\omega t + \frac{\pi}{2} \right]) + \frac{1}{2} \mu r^2 \\ \nonumber
\mathrm{D}_{yy}' & = & \frac{1}{2} \mu r^2 - \frac{3}{2} \mu r^2 
\cos (2 \left[ \omega t + \frac{\pi}{2} \right]) \\ \nonumber
\mathrm{D}_{xy}' & = & \frac{3}{2} \mu r^2 \sin (2 \left[ 
\omega t + \frac{\pi}{2} \right]) \\ \nonumber
\mathrm{D}_{zz}' & = & - \mu r^2.
\end{eqnarray}
It is easy to see that 
$\dot{\mathrm{D}}_{xx}' = - \dot{\mathrm{D}}_{xx}$, 
$\dot{\mathrm{D}}_{yy}' = - \dot{\mathrm{D}}_{yy}$, 
$\dot{\mathrm{D}}_{xy}' = - \dot{\mathrm{D}}_{xy}$ and 
$\dot{\mathrm{D}}_{zz}' = \dot{\mathrm{D}}_{zz} = 0$.  
That is, the time derivative
(to all orders) of the magnitude of the
moment of inertia tensor is identically zero; there
is {\it no} gravitational radiation.  Baker's formula, Eq.\,(\ref{Baker}),
however, predicts double the radiation of the two-body situation.

Next, consider two bodies of mass $m_1$ and $m_2$ constrained
to move along a straight line, which we will identify with the $z$-axis.
We fix the centre of our coordinate system at their centre of mass.
Their distance apart is $r$.  The components of the moment of inertia
tensor are
\begin{eqnarray}
\mathrm{D}_{xx} & = & - \mu r^2 \\ \nonumber
\mathrm{D}_{yy} & = & - \mu r^2 \\ \nonumber
\mathrm{D}_{zz} & = & 2 \mu r^2
\end{eqnarray}
with $\mu$ the reduced mass, as before.  
(By allowing one of the masses
to be much larger than the other we can use these formulae to
analyse the motion of a single body.)  From these we calculate
\begin{equation}
\left| \dddot{\mathrm{D}}_{\alpha \beta} \right|^2 =
24 \mu^2 \left( r \dddot{r} + 2 \ddot{r} \dot{r} \right)^2
\end{equation}
or, to compare with Baker's formula,
\begin{equation}
\left| \dddot{\mathrm{D}}_{\alpha \beta} \right|^2 =
24 \left( r \dot{f} + 2 \dot{r} f \right)^2.
\label{rectilinear}
\end{equation}
If the force (a sort of reduced force, acting on the reduced mass) is
constant, Baker's formula gives no gravitational radiation; but 
Eq.\,(\ref{rectilinear}) shows that there is still some given off.

Baker's formula has thus been shown to be completely in error both ways,
in predicting gravitational radiation when there is none, and predicting
none when there is some.  Thus it cannot be used even as a rough 
guide.  The examples given are only slightly different from those
in Baker (2006) and would be a reasonable first approximation
to some astronomical objects 
(orbits of more than two objects are fairly common, as are linear
jets), so his assertion in that paper of the usefulness of his
formula in astrophysics is unfounded.

Such a conclusion would not appear to have a big impact on the
gravitational-wave community, since Baker's formula has not been
used in astrophysical circles (where more accurate techniques are
customary).  It has been employed in another context, however,
and that forms our third example.

\section{Laboratory gravitational waves?}

Baker, Li \& Li (2006) describe an apparatus intended to generate
and detect high-frequency gravitational waves in a laboratory.  Only
the generation side concerns us here.

Two targets, 20m apart, are hit with high-intensity laser pulses,
directed such that they are momentarily accelerated in opposite
directions; the authors consider them to emulate, for the duration
of the pulse, a two-body orbiting system.  The 23TW pulses last for
33.9fs and are repeated ten times per second.  Using Eq.\,(\ref{Baker})
the authors calculate that they generate $5.5 \times 10^{-15}$
watts of gravitational radiation\footnote{The authors tacitly assume
that the pulse changes intensity linearly over its duration, inserting
the calculated radiation-pressure force for 23TW and the duration
of 33.9 fs in Baker's formula.  It might be more accurate to take the
pulse as a square wave, staying at something like its peak intensity
for the duration.  In principle, the difference is important since
the factors enter as squares, and the average of a square is not the
square of the average.  When I calculate some numbers I will look at
both possibilities.}.

There is one major problem with this analysis.  Baker et al.\, apply their
formula for the duration of one laser pulse, $3.39 \times 10^{-14}$s.
Light travels only 10.6$\mu$m in this time (something they mention, but
without any apparent consideration of its implications).  To analyse bodies
not in causal contact as if they were part of a Newtonian object, moving
in Newtonian ways, is a very questionable procedure.
Indeed, it is not clear that the essentially
classical definition of a moment of inertia tensor and its time derivatives
can be made relativistically meaningful, nor that it would retain its
role as a source of gravitational radiation if it were.  And it is
certainly not justified to apply Eq.\,(\ref{power}), which is averaged over
a complete period of the gravitational waves, to a tiny fraction of an
orbit.  By looking at such a short pulse Baker et al.\, (2006) have 
removed themselves from the assumptions underlying the starting point of
Baker's (2006) derivation, Eq.\,(\ref{power})
and so have no grounds for believing their
resulting numbers.

As far as one can apply Eq.\,(\ref{power}) to the apparatus of Baker et al.\,
one must be
restricted to a region in causal contact, a single laser target.  Indeed,
not all of that: ordinary matter is not rigid on this time scale.  The
impact of the laser pulse on the rear of a target cannot be known at the
front until a sound wave can traverse the intervening distance.  If this
period is much shorter than the pulse, the whole object cannot accelerate
(as Baker et al.\, tacitly assume); instead a sound wave is set
ringing through the target.

To clarify the picture it is useful to have some numbers.  Since Baker
et al.\, (2006) give no details about their laser targets, I will use
some nominal values (exact figures are not important here, as will be
clear).  For a nominal sound speed in steel of 5790 m s$^{-1}$, the
pulse of $3.39 \times 10^{-14}$s has penetrated a distance of
$1.96 \times 10^{-10}$m by the time it ends: a layer of atomic
dimensions\footnote{A true analysis of this interaction of radiation with
matter should, of course, be done in a quantum context.}.
Assuming a laser spot size similar
to their quoted detector laser, $1.96 \times 10^{-5}$m$^2$ (not all of
which is in causal contact sideways!), and a density of steel of 
$7.9 \times 10^3$kg m$^{-3}$, we have an accelerated mass of something like
$3.0 \times 10^{-11}$kg.

We now turn to Eq.\,(\ref{rectilinear}) to calculate the gravitational
radiation of this accelerated mass.  First using the Baker et al.\,
assumption of a constant change in force, 

\begin{equation}
\left| \dddot{\mathrm{D}}_{\alpha \beta} \right|^2 =
\frac{98}{3} \frac{\dot{f}^4 t^6}{\mu^2}
\end{equation}
which gives a radiated power of $2.5 \times 10^{-39}$W.  If, on the other
hand, we assume a constant force,
\begin{equation}
\left| \dddot{\mathrm{D}}_{\alpha \beta} \right|^2 =
96 \frac{f^4 t^2}{\mu^2}
\end{equation}
resulting in $7.3 \times 10^{-39}$W.  The difference between these numbers
and the calculation in Baker et al.\, (2006) amounts to twenty-seven
orders of magnitude.  This difference may in principle arise either from
the simplifications of Baker (2006) or from the possibility that 
Eq.\,(\ref{power}) is simply inappropriate for very short periods of time;
in either case, Eq.\,(\ref{Baker}) must be discarded.

\section{Conclusions}

A simplified formula for a relativistic effect has been shown to be
completely unreliable, giving infinitely wrong answers in two instances,
and an error of something like twenty-seven orders of magnitude in
a third.  The latter number is not firm, since there is some question as
to whether its own basis is justified; but it is certain that
the formula of Baker (2006) cannot be used.  Gravitational waves are
not to be generated in the laboratory in the forseeable future.

The lesson of this episode is 
that simplified formulae must be justified and carefully
handled, since it is quite possible to push them beyond their applicability.
In addition, Newtonian approximations of relativistic effects must be
carefully examined for tacit assumptions that make their results untenable.
In particular, short time scales (as well as speeds comparable to light)
make Newtonian expressions unreliable.

\end{document}